\documentclass[reprint,aps,prb,superscriptaddress,nofootinbib,longbibliography]{revtex4-1}
\usepackage{braket}
\usepackage{times} 
\usepackage{xcolor}
\usepackage{amsfonts,amssymb,amsmath}
\usepackage[T1]{fontenc}

\usepackage{qcircuit}
\usepackage{xr-hyper}
\usepackage{hyperref}
\usepackage{enumitem}
\usepackage{comment}
\usepackage{makecell}
\usepackage[capitalize]{cleveref}
\usepackage{booktabs}
\usepackage{hhline}

\makeatletter
\newcommand*{\addFileDependency}[1]{% argument=file name and extension
  \typeout{(#1)}
  \@addtofilelist{#1}
  \IfFileExists{#1}{}{\typeout{No file #1.}}
}
\makeatother

\usepackage{amsthm}

\newcommand*{\myexternaldocument}[1]{%
    \externaldocument{#1}%
    \addFileDependency{#1.tex}%
    \addFileDependency{#1.aux}%
}
\myexternaldocument{sup}
\hypersetup{colorlinks=true,citecolor=blue,linkcolor=blue, urlcolor=blue}
\hypersetup{linktocpage}
\newcolumntype{M}[1]{>{\centering\arraybackslash}m{#1}}
\newcolumntype{N}{@{}m{0pt}@{}}
\usepackage{environ}

\bibpunct{[}{]}{,}{n}{}{}

\usepackage{tikz}
\usetikzlibrary{arrows,snakes,backgrounds}

\myexternaldocument{sup}

\NewEnviron{eqs}{%
\begin{equation}\begin{split}
    \BODY
\end{split}\end{equation}
}

\begin{document}

\title{Beyond Boson Sampling: Higher Spin Sampling as a Practical Path to Quantum Supremacy}

\author{Chon-Fai Kam}
\affiliation{Department of Physics, University at Buffalo, SUNY, Buffalo, New York 14260, USA}

\author{En-Jui Kuo}
\affiliation{
Department of Electrophysics, National Yang Ming Chiao Tung University, Hsinchu, Taiwan, R.O.C.}

\begin{abstract}
Since the dawn of quantum computation science, a range of quantum algorithms have been proposed, yet few have experimentally demonstrated a definitive quantum advantage. Shor’s algorithm, while renowned, has not been realized at a scale to outperform classical methods. In contrast, Fock-state boson sampling has been theoretically established as a means of achieving quantum supremacy. However, most existing experimental realizations of boson sampling to date have been based on Gaussian boson sampling, in which the input states consist of squeezed states of light. In this work, we first introduce spin sampling for arbitrary spin-$S$ states as a practical path to quantum supremacy. We identified a quasi-linear scaling relation between the number of sites $m$ and the number of spins $n$ as $m\sim n^{1+\epsilon}$, where $\epsilon=3/(2S)$ is inversely proportional to the spin quantum number. This suggests that, within a spin system, an equivalent Fock-state boson sampling task in the linear mode region, characterized by $m\sim \Omega(n^{1+\epsilon})$, is experimentally feasible with reduced resource requirements.
\end{abstract}

\maketitle

\section{Introduction}
One of the central goals in quantum computation science is to design algorithms that outperform their classical counterparts. A prominent example is Shor’s algorithm for integer factorization \cite{ekert1996quantum}, which theoretically offers an exponential speedup over the best-known classical methods \cite{lenstra1993number}. However, experimental demonstrations of Shor’s algorithm have been limited to factoring small numbers—like 15 \cite{vandersypen2001experimental, lucero2012computing, lu2007demonstration} and 21 \cite{martin2012experimental}—due to the complexity of the required quantum circuits (involving components such as the quantum Fourier transform \cite{nielsen2002quantum, shor1997polynomial} and modular exponentiation \cite{vedral1996quantum}) and current hardware constraints. For instance, research suggests that the number of CNOT gates required to factor an $n$-bit integer using Shor’s algorithm is approximately $217 n^3 \log^2(n) + 4n^2 + n$ \cite{liu2021cnot}. In the case of factoring a 10-bit number, this would amount to roughly one million gates—highlighting the significant computational resources required for practical implementation. As a result, these early proof-of-concept experiments factor numbers that classical computers can easily handle in microseconds, leaving the practical quantum advantage of Shor’s algorithm still on the horizon.

In contrast, boson sampling has emerged as an alternative approach aimed at demonstrating quantum supremacy \cite{arute2019quantum}. Proposed by Aaronson and Arkhipov \cite{aaronson2011computational} in 2012, boson sampling employs specialized photonic systems that harness the behavior of linear optics to sample from the distribution of bosonic states. Here, the probability of observing a specific output is linked to the squared absolute value of a matrix permanent—a function known to be $\#$P-hard \cite{valiant1979complexity,scheel2004permanents}. Valiant's theorem \cite{valiant1979complexity} proves that computing the permanent for even simple 0-1 matrices is $\#$P-complete, underscoring the substantial computational challenge it poses for classical simulation. The key complexity-theoretic implication is this: if a classical computer could efficiently simulate boson sampling, it would imply that a $\#$P-hard problem can be solved efficiently in polynomial time \cite{aaronson2011computational,brod2020review}. Aaronson and Arkhipov showed that such an efficient classical simulation would lead to the collapse of the polynomial hierarchy to at least the third level.

The original Aaronson-Arkhipov boson sampling proposal assumes ideal conditions, such as precise single-photon inputs, which are experimentally challenging \cite{shchesnovich2014sufficient}. The ideal implementation of boson sampling requires deterministic single-photon sources, low-loss linear interferometers, and high-efficiency detectors \cite{brod2019photonic}. Early experiments relied on probabilistic sources, particularly spontaneous parametric down-conversion (SPDC), which generates photon pairs probabilistically \cite{broome2013photonic, tillmann2013experimental}. By detecting one photon, the other is heralded as a single photon. However, this method necessitates post-selection to ensure the correct photon count, becoming exponentially inefficient as the number of photons increases. Achieving quantum advantage requires scaling boson sampling to large photon numbers, typically 30 or more \cite{broome2013photonic}. However, probabilistic sources present challenges in generating an $n$-photon state, as the probability declines exponentially with increasing $n$ \cite{meyer2022scalable}. Additionally, losses in the interferometer and inefficiencies in detection further reduce experimental fidelity \cite{shi2022effect}. Alternatively, weak coherent states generated by attenuated laser pulses, with a mean photon number of one, follow Poisson statistics and may include multiple photons, introducing multi-photon contamination in the output \cite{bressanini2023gaussian, chabaud2021efficient}. 

Nevertheless, in 2020, a research team from China demonstrated Gaussian boson sampling (GBS) using their photonic quantum computer, Jiuzhang, with 50 squeezed states and detecting up to 76 photons \cite{zhong2021phase}. Jiuzhang completed the GBS task in approximately 200 seconds (roughly 3.3 minutes) \cite{zhong2021phase}. The researchers estimated that Fugaku, a leading classical supercomputer at the time, would take on the order of 600 million years to simulate the same GBS task with comparable fidelity \cite{bulmer2022boundary}, highlighting a quantum computational advantage for this specific problem. One year earlier, in 2019, Google’s research team in the United States announced that they had achieved quantum supremacy using their 53-qubit superconducting quantum processor, Sycamore \cite{arute2019quantum}. They performed random circuit sampling (RCS), a task designed to be intractable for classical computers \cite{boixo2018characterizing,aaronson2016complexity}. Sycamore completed the RCS task in approximately 200 seconds, while the team estimated that Summit, a leading classical supercomputer, would take around 10,000 years to perform the equivalent computation \cite{arute2019quantum}. Although distinct from Gaussian boson sampling, random circuit sampling similarly showcases quantum advantage by leveraging the complexity of quantum circuits \cite{aaronson2016complexity,zhong2020quantum}.

Random circuit sampling involves executing a random quantum circuit and sampling its output probability distribution. Its classical intractability arises from the exponential size of the Hilbert space for $n$ qubits, which makes simulating the output distribution extremely resource-intensive. Unlike Gaussian boson sampling, whose hardness is directly tied to \#P-hard problems like computing matrix permanents, RCS’s intractability relies on conjectured difficulties in quantum complexity theory, such as the inability of classical computers to efficiently replicate the output statistics of deep random circuits \cite{aaronson2016complexity}. Recent research introduced improved classical algorithms that significantly reduced the estimated classical simulation time, leading to ongoing discussions about the extent of quantum supremacy. A recent study demonstrated that using 1432 GPUs, researchers were able to perform classical simulations seven times faster than Sycamore's original experiment \cite{pan2021solving}. Additionally, further improvements have reportedly achieved a 50-fold speedup, suggesting that classical methods are rapidly closing the gap \cite{liu2023closing}.

Intriguingly, quantum computing platforms extend beyond photonic and superconducting qubits, with spin qubits serving as a promising alternative. Formed from electron or hole spins in semiconductor quantum dots, spin qubits offer potential for scalability and efficiency. Their close integration with semiconductor manufacturing facilitates compact designs and exceptional scalability. These advantages could reduce their footprint from bulky superconducting systems to portable, laptop-sized devices \cite{kam2024submicrosecond, kam2024fast}. As discussed, a clear demonstration of quantum supremacy remains an experimental challenge, making it imperative to explore different experimental platforms. Boson sampling, which relies on calculating permanents, is computationally intensive for classical computers. In contrast, fermionic sampling, based on determinants, can be efficiently solved using Gaussian elimination. An intriguing question in this context is whether spin-based sampling, whose computational complexity relatively unexplored, could pave the way for novel quantum advantage protocols. In this work, we demonstrate that spin-sampling can exhibit quantum advantage, analogous to its bosonic counterpart.

In the literature, spin-sampling has been proposed to map the boson sampling problem onto the sampling problem for long-range spin-1/2 $XY$ Hamiltonian \cite{peropadre2017equivalence,kuo2022boson}. This mapping is rigorous for hard-core bosons under the dilute limit, $n\ll m (m \sim \Omega(n^4)$  , where $n$ represents the number of bosons, and $m$ denotes the number of modes. Consequently, the hardness arguments from boson sampling can be applied to spin systems, suggesting that simulating spin-sampling could be just as computationally complex as boson sampling \cite{peropadre2017equivalence,kuo2022boson}. However, conducting a spin-sampling experiment presents substantial challenges. Fully leveraging these complexity arguments necessitates a large-scale system of spin-1/2 particles, but scaling up quantum systems introduces noise and control difficulties, making implementation highly nontrivial. Consequently, as of now, no experimental realization of spin-sampling has been achieved in a physical quantum system.

More precisely, in their foundational proof on the computational hardness of boson sampling \cite{aaronson2011computational}, Aaronson and Arkhipov established a key condition requiring the mode number $m$ to scale as $\Omega(n^2)$, where $n$ represents the number of bosons—a necessity arising from the birthday paradox \cite{sayrafiezadeh1994birthday}. This is further reinforced by the assumption that small sub-matrices of a Haar-random matrix follow an i.i.d. Gaussian distribution \cite{petz2004asymptotics}. This high-mode condition is integrated into all existing spin-$1/2$ boson sampling proposals via the boson-spin mapping \cite{peropadre2017equivalence, kuo2022boson}, where a quartic scaling $m = \Omega(n^4)$ is imposed to ensure that the spin sampling distribution remains indistinguishable from the standard boson sampling distribution. This ultra-high mode condition seems inefficient, as the quartic scaling inherently includes the quadratic scaling. This suggests that requiring a higher mode count may reduce the practical utility of spin-based boson sampling compared to the original bosonic framework. If the additional modes don’t contribute meaningfully to computational advantage or experimental feasibility, then the trade-off between complexity and usefulness becomes a concern. 

This critical issue remained unsettled until recently, when a theoretical study \cite{bouland2023complexity} demonstrated that even in the low-mode region, where $m\sim 2.1n$, quantum advantages persist in Boson sampling despite the presence of high collision pairs and correlated Gaussian distributions. Notably, no classical algorithm can reproduce boson sampling under the same strong computational hardness assumptions as those in the high-mode region \cite{aaronson2011computational, spring2013boson,lund2014boson, chakhmakhchyan2017boson}, where the mode number scales as $m\sim n^2$. Leveraging the hardness result in the linear mode regime of boson sampling paves the way for our efficient higher spin sampling protocol. 

Specifically, in this work, we introduce a spin-sampling task for general spin-$S$ systems, where $S$ is an arbitrary half-integer. We build upon the original boson sampling problem, in which the occupation number per mode can reach infinity. In our case, the target Hilbert space is the spin subspace, where each site hosts spin excitations constrained to be at most $2S$. The full occupancy of all available levels at a single site is considered outside the target space. Our theoretical findings indicate that assuming the original boson sampling is computationally hard for large $n$ and $m$, subject to the quasi- or super-linear scaling $m \sim n^{1+p}$, where $p$ is any positive real number, then the corresponding spin-sampling problem is also computationally hard for a system of $n$ spin-$S$ particles and $m$ sites, following the scaling $m\sim n^{1+3/(2S)}$, which is quasi-linear for higher spin. When the spin quantum number $S$ is moderate, such as $S=6$, our spin-sampling protocol requires only $m\sim n^{1.25}$ to achieve computational hardness. This demonstrates a significant reduction in resource requirements, making our higher spin sampling far more efficient. Compared to \cite{peropadre2017equivalence}, our approach achieves an exponential improvement in mode number $m$, reducing waste and enhancing experimental feasibility.

A further motivation for developing higher spin sampling is that all current boson sampling experiments use Gaussian boson sampling, where the input light consists of squeezed coherent states \cite{kam2023coherent} rather than Fock states. However, recent developments \cite{grier2022complexity, oh2022classical, oh2024classical, yang2024speeding,ehrenberg2025transition} have shown that in the low-mode regime—where the number of modes scales quasi-linearly with the number of bosons—an efficient classical algorithm exists for solving noisy Gaussian boson sampling. Based on these observations, Gaussian boson sampling requires a greater number of modes compared to Fock-state boson sampling. While Bouland et al. \cite{bouland2023complexity} asserts that Gaussian boson sampling can also be computationally hard in the linear regime, the existence of efficient classical algorithms suggests that the original Fock-state boson sampling is likely more challenging than Gaussian boson sampling. There are ongoing experimental efforts to directly implement the Fock-state boson sampling protocol, where $m$ scales as $\Omega(n)$. However, as we have emphasized, single-photon sources are challenging to prepare. From this perspective, in addition to Fock-state boson sampling, our higher spin sampling protocol provides a compelling approach to achieving quantum supremacy in the more experimentally accessible low-mode region, where $m$ scales quasi-linearly with $n$. Our findings could unlock new avenues for the practical implementation of quantum algorithms, offering computational advantages in near-term quantum experiments and enhancing feasibility for experimental realizations.

\section{Methods} 
In the original Fock-state Boson sampling, when restricting to the high-mode region where the number of modes $m$ is much larger than the boson number $n$, each mode predominantly contains either zero or one boson. This configuration enforces a hard-core boson behavior, as only a single boson can occupy each mode, effectively mirroring the Pauli exclusion principle. This scenario allows for a natural mapping to a spin-$\frac{1}{2}$ system: $|\downarrow\rangle \mapsto |0\rangle$ and $|\uparrow\rangle \mapsto |1\rangle$, as well as a mapping between the Heisenberg algebra $h_4$ generators and the SU$(2)$ algebra generators: $\hat{a}^\dagger \mapsto \hat{\sigma}^+$, $\hat{a} \mapsto \hat{\sigma}^-$ and $\hat{n}\mapsto (\hat{\sigma}^z + 1)/2$. Under this transformation, the presence of a boson in a given mode corresponds to the spin-up state, while an empty mode corresponds to the spin-down state. Thus, the general linear Hamiltonian that couples the input and output modes is uniquely transformed into an equivalent linear Hamiltonian that couples the input and output spins.

Unfortunately, in the general spin-$S$ representation, a direct linear mapping between the Heisenberg algebra generators and the SU$(2)$ algebra generators does not exist. The well-known Holstein-Primakoff representation introduces an additional $\sqrt{2S - \hat{S}^z}$ factor, The generators of the Heisenberg algebra $h_4$ and SU$(2)$ algebra transform as: $\hat{a}^\dagger \mapsto \sqrt{2S - \hat{S}^z} S^+$, $\hat{a} \mapsto S^- \sqrt{2S - \hat{S}^z}$ and $\hat{n} \mapsto \hat{S}^z + S$. Although this mapping translates the soft-core boson dynamics into spin dynamics, the corresponding spin Hamiltonian is no longer linear in the SU$(2)$ algebra generators. This poses challenges for experimental implementation, except in the large-spin limit.

In the following, we adopt an alternative approach. We begin with a linear Hamiltonian that is expressed in terms of the generators of a broad class of generalized bosonic algebras. To proceed, let us recall the bosonic commutation relations governing the annihilation and creation operators, denoted by $a_i$ and $a_j^{\dagger}$, for given modes $i$ and $j$ are $[a_i, a_j^{\dagger}] = \delta_{ij}$ and $[a_i^{\dagger}, a_j^{\dagger}] = [a_i, a_j] = 0$. For generalized bosons, the final two commutation relations remain unaltered, whereas the first commutation relation is adjusted by a characteristic function $F(n)$ that depends on the single-mode occupation number $n$. In particular, the first commutation relation, has been modified as follows
\begin{equation}
    [a_i, a_j^{\dagger}] = \delta_{ij} \sum_{n_i=0}^{\infty} F(n_i) |n_i\rangle \langle n_i|.
\end{equation}
Here, the generalized Fock state $|n_i\rangle$ with occupation number $n_i$ in mode $i$ is defined as $|n_i\rangle\equiv \frac{1}{f(n_i)}(a_i^{\dagger})^{n_i} |0\rangle$, where $f(n)$ is another characteristic function that defines the generalized boson. The characteristic functions $f(n)$ and $F(n)$ are interconnected, and their precise relationship depends upon specific physical systems. For example, in a standard bosonic system, one has $F(n)=1$ and $f(n)=\sqrt{n!}$, whereas in the standard spin-$S$ system, one has $F(n)=(n-2S)\theta(2S-n)$ and $f(n)=\sqrt{n!(2S)!/(2S-n)!}$. As such, a multi-mode Fock state fro the generalized bosons with $m$ modes can be written as
\begin{equation}
    |n_1, n_2, \dots, n_m\rangle = \left(\prod_{i=1}^{m} \frac{1}{f(n_i)}\right) a_1^{\dagger n_1}a_2^{\dagger n_2}\cdots a_m^{\dagger n_m} |0\rangle.
\end{equation}
The Aaronson-Arkhipov Fock-state boson sampling protocol involves sending a Fock state containing $n$ photons across $m$ mode into a randomly selected linear mode-mixing circuit, represented by a unitary matrix $U$. This process enables mode-mixing, wherein the linear optical network \( U \) transforms the input mode operators \( \{a_i\}_{i=1}^{m} \) into corresponding output operators $b_i = \sum_{j=1}^{m} U_{ij} a_j$. The probability of observing the outcome $\mathbf{k} = (k_1, \dots, k_m)$ when the input configuration is $\mathbf{l}= (l_1, \dots, l_m)$, where $\sum_i k_i = \sum_i l_i = n$, is determined by
\begin{equation}
    \mbox{Pr}(\mathbf{k}|\mathbf{l}) = \frac{\left| \text{Perm}(\Lambda[\mathbf{k}|\mathbf{l}]) \right|^2}{(\prod_i l_i!)( \prod_i k_i!)}.
\end{equation}
Here, the permanent of an $N \times N$ matrix \( A = (A_{i,j})_{i,j} \) defined similarly to the matrix determinant, but differs in that all signs in the summation remain positive
\begin{equation}
\text{Perm}(A) 
\equiv \sum_{\sigma \in S_N} \prod_{i=1}^{N} A_{i, \sigma(i)},
\end{equation}
where $S_N$ denotes the symmetric group on $N$ elements. Additionally, $U[\mathbf{m}|\textbf{n}]$ is an $N \times N$ matrix constructed by repeating the \( i \)-th column of \( U \) \( l_i \) times, and the \( j \)-th row \( k_i \) times.

\section{Scaling Relation}
This section analyzes how the error bound for distributions beyond the target spin subspace where spin excitations are restricted to at most $2S$, scales with the number of spins $n$, the number of sites $m$, and a fixed spin quantum number $S$. The method employed has been used in Ref. \cite{peropadre2017equivalence} in the context of spin-$\frac{1}{2}$ sampling. However, it has never been extended to the general spin-$S$ sampling. In particular, we demonstrate that for a fixed error, the scaling relation between $n$ and $m$ in spin-$S$ sampling is quasi-linear, characterized by $m \sim n^{1 + \epsilon}$, where $\epsilon \equiv \frac{3}{2S}$ is inversely proportional to the spin quantum number $S$. Thus, the number of cites required to achieve quantum supremacy is significantly lower than that of its spin-$1/2$ counterpart, which involves a quartic scaling. This constitutes the central discovery of this paper.

In particular, the proof presented in \cite{peropadre2017equivalence} relies on projecting the bosonic Hilbert space onto the hardcore boson subspace. For the general spin-$S$ setting, we perform a similar analysis by projecting onto the spin subspace, where spin excitations are restricted to at most $2S$. Let us examine the following Hamiltonian
\begin{equation}
H_{S} = \sum_{i,j=1}^m( \hat{S}_{j,\text{out}}^{x\dagger} R_{ji} \hat{S}^x_{i,\text{in}} + \text{H.c.})
+ B\sum_{j=1}^m(\hat{S}^z_{j,\text{out}}+ \hat{S}^z_{j,\text{in}})
\label{eq:harmonic}
\end{equation}
which is an Ising Hamiltonian couples the input and output subsets in a spin sampling task, while a strong magnetic field enforces an effective non-local $XY$ Hamiltonian on the spin system. The initial state $\ket{\phi(0)} = \hat{S}_{1,\text{in}}^{+} \cdots \hat{S}_{N,\text{in}}^
+\ket{\mathrm{vac}}$ evolves into a superposition state characterized by $n$-spin sampling superposition at a later time $t=\pi/2$
\begin{equation}
\ket{\phi(\pi/2)} = (-i)^n\prod_{i=1}^N \sum_j R^*_{ji} \hat{S}_{j,\text{out}}^+\ket{\mathrm{vac}}
\label{eq:bs-state}
\end{equation}
with a spin distribution given by the permanent discussed earlier. We evaluate the bound of bunching events by projecting the $n$-spin superposition state $|\phi\rangle$ onto the target spin subspace. This projection can be expressed as $\ket{\phi} = Q\ket{\phi} + \ket{\varepsilon}$, where $Q\ket{\phi}$ represents the component within the target subspace, and $\ket{\varepsilon}$ encompasses bunched states where at least one site exhibits a spin excitation of $2S+1$. To ensure the elimination of errors $\ket{\varepsilon}$ in post-selection while preserving sampling efficiency, the number of cites must exceed the number of excitations.

Consider a spin state \( |\psi\rangle \) that initially matches the starting distribution of the spin sampling problem, such that \( |\psi(0)\rangle = |\phi(0)\rangle \). The evolution of this state under the spin model is governed by $i\partial_t |\psi\rangle = Q H_{S} Q |\psi\rangle$. To quantify the sampling error introduced by working with spins, we define the error state as $|\delta\rangle = Q|\phi\rangle - |\psi\rangle$. This formulation separates the projected target state from its evolution within the spin system, allowing for a precise analysis of sampling errors and deviations from the ideal state. The solution to this error is as follows
\begin{equation}
\ket{\delta(t)} = -i\int_0^t e^{-i QH_{S}Q (t-\tau)} Q H_{S} \ket{\varepsilon(\tau)} \mathrm{d}\tau.
\label{eq:delta-equation}
\end{equation}

\begin{figure}
    \centering
    \includegraphics[width=\columnwidth]{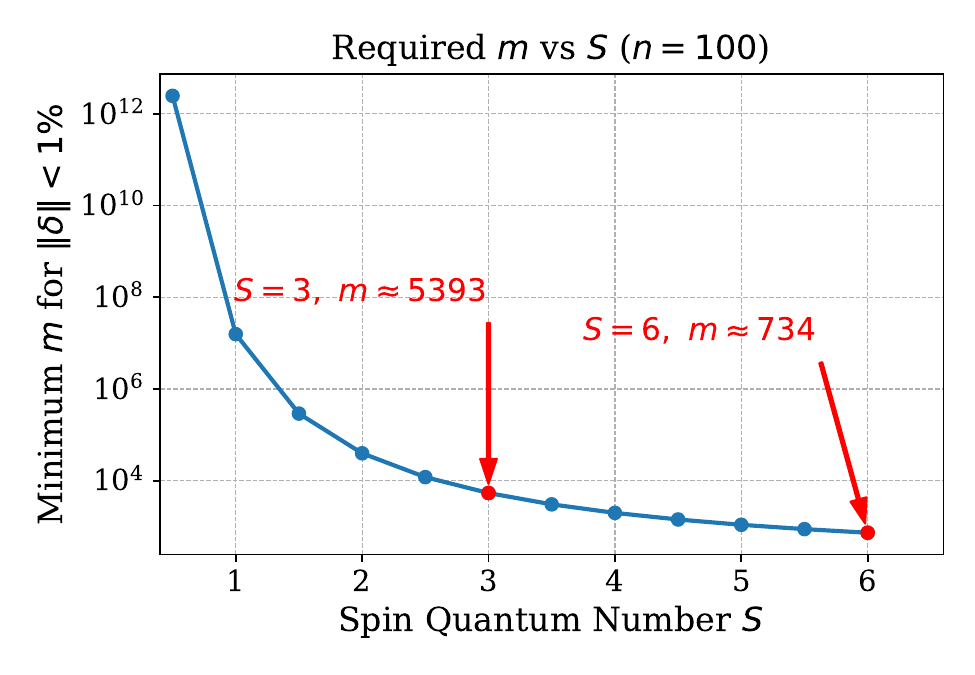}
    \caption{The scaling relationship between the number of spins $n$ and the number of sites $m$ for various spin quantum numbers $S$.}
    \label{fig:spin}
\end{figure}

We now bound the maximum error probability as an integral of two norms. For that, we realize that out of $\varepsilon$, $Q H_{S}$ cancels all terms that have more than $2S$ output site with double occupation. Thus, 
\begin{equation}
\|\delta\| \leq \int_0^t \|Q H_{S} P\|\cdot \|P\ket{\varepsilon(\tau)}\| \mathrm{d}\tau,
\label{eq:spin-error}
\end{equation}
Here, given an operator $A$, its operator norm is defined as $\|A\|\equiv \sup_{\|x\|=1}\|Ax\|$, which represents the largest singular value of $A$. In the above, $P$ is a projector onto the states with at most $2S+1$ spins on the same site of output site. The value $\|P\ket{\varepsilon(\tau)}\|^2=\|P\ket{\phi(\tau)}\|^2$ is the probability of finding a single bunched event in the full Hilbert state. We can use the spin-$S$ birthday paradox formula Eq.~\eqref{eq:spin-birth} to evaluate the above probability
\begin{equation}\label{SM_eq:paradox_bound}
\|P\ket{\varepsilon(\tau)}\| \leq \mathcal{O}\left(\frac{n^{S+\frac{1}{2}}}{m^{S}}\right),
\end{equation}
We now can use results from \cite{peropadre2017equivalence} and bound the operator norm $\|Q H_{S}P\|$ as strictly smaller than the maximum kinetic energy of $n$ bosons in the original model, so that
\begin{equation}\label{SM_eq:kinetic_bound}
\|Q H_{S} P\| \leq n.
\end{equation}
Combining bounds shown in Eq.~\eqref{SM_eq:paradox_bound} and Eq.~\eqref{SM_eq:kinetic_bound}, we finally find
\begin{equation}
\|\delta(t)\| \leq t \times \mathcal{O}\left(\frac{n^{S+\frac{3}{2}}}{m^{S}}\right).
\label{equation_finalbound}
\end{equation}
In particular, for the spin-1/2 case, we recover the error bounds found in Ref.~\cite{peropadre2017equivalence}. In Fig \ref{fig:spin}, we present the number of sites required to achieve quantum supremacy for different spin quantum numbers $S$. One can clearly observe that the number of sites is exponentially reduced compared to the spin-1/2 case, and remains approximately of the same order as the number of spins, even for an experimentally achievable spin quantum number $S=6$.

\section{Experimental Realization}
The major experimental platforms for performing spin sampling include, but are not limited to, trapped ion qubits, spin qubits, the D-Wave platform, and cold atom systems. Each of these platforms has its own advantages and limitations. For example, trapped ion qubits, controlled through laser-driven interactions, are particularly well-suited for spin sampling tasks due to their high degree of controllability, enabling the engineering of long-range interactions. These systems support programmable spin models and the encoding of higher spin states in qudits, allowing for simulations beyond conventional spin-$1/2$ models. The technology for spin-$1/2$ systems is more advanced. Recent experiments have successfully simulated systems with up to 512 ions, representing a major breakthrough \cite{guo2024site}. For higher spin states, researchers have demonstrated simulations with $S = 3$ using up to 8 ions, leveraging qudit-based approaches \cite{monroe2021programmable}. Scaling to larger system sizes for higher spins presents challenges due to increased control complexity, but reviews highlight promising prospects for surpassing 100 qubits in future developments \cite{foss2024progress}.

D-Wave's quantum annealers employ superconducting qubits to solve optimization problems formulated as Ising models. Hence, they can naturally be used for spin-sampling tasks involving large-scale Ising models with strong external magnetic fields. The Advantage system contains over 5,000 qubits with programmable couplings, enabling long-range interactions \cite{willsch2022benchmarking}. Introduced in 2022, the Advantage2 system maintains a similar scale, with plans to exceed 7,000 qubits for enhanced large-scale simulations \cite{vert2024benchmarking}. However, these systems exclusively utilize spin-$1/2$ qubits, limiting their applicability for higher-spin sampling tasks. 

Cold atom systems, particularly those utilizing ultracold atoms in optical lattices or tweezers, provide significant capabilities for spin sampling. Alkaline-earth atoms such as strontium-87, with a nuclear spin of $I=9/2$, enable the simulation of spin models with $S$ up to $9/2$ in lattice-based experiments \cite{browaeys2020many}. Optical lattices can accommodate thousands of atoms, but interactions are typically limited to nearest-neighbor coupling \cite{gross2017quantum}. To address this constraint, long-range interactions can be engineered using techniques such as Rydberg dressing or dipole-dipole interactions, as demonstrated in studies with Rydberg atoms \cite{scholl2021quantum}. While these methods extend to higher-spin models, current experimental efforts still primarily focus on spin-$1/2$ systems.

Spin qubit systems have progressed relatively slowly due to the weak interactions between qubits, which limit scalability and gate fidelity. Currently, mainstream spin qubits are based on spin-$1/2$ electron systems, with the simplest implementation involving a single electron confined in a quantum dot. This approach utilizes the intrinsic spin degree of freedom of electrons, which can be manipulated using external magnetic or electric fields to perform quantum operations. Singlet-triplet qubits, formed by two electrons, create a two-level system with total spin states $S=0$ and $S=1$ \cite{burkard2023semiconductor}. Since these states are effectively binary, they do not extend $S$ beyond $1/2$ per site. The largest reported spin qubit array consists of 12 qubits, fabricated on a 300 mm semiconductor platform \cite{george202412}. This represents the current state-of-the-art for scaling. However, interactions in spin qubits are typically limited to nearest-neighbor coupling. Engineering long-range interactions requires coupling through microwave cavities, which are not yet standard for large arrays \cite{burkard2023semiconductor}. Consequently, spin qubits are not well-suited for spin sampling tasks unless significant modifications are made.

The comparison of experimental platforms for implementing higher-spin sampling tasks is summarized in Table~\ref{table1}. Among existing platforms, trapped ion systems offer the best balance between achieving high spin values $S$, while maintaining large system sizes $n$. Cold atom systems also present promising potential, particularly if the challenge of long-range interactions can be overcome using Rydberg techniques. In contrast, D-Wave systems and spin qubits are less suited for pursuing quantum advantage in higher-spin sampling due to their inherent limitation of supporting only spin-$1/2$ per site.

\begin{table*}\label{table1}
\centering
\begin{tabular}{lcccc}
\toprule
\textbf{Platform} & \textbf{Maximum $S$} & \textbf{Maximum $n$} & \textbf{Long-Range Interactions} & \textbf{Suitability for XY/Ising Models} \\
\midrule
Trapped Ions & 3 & 512 ($S=\frac{1}{2}$) \cite{guo2024site}, 8 ($S=3$) \cite{monroe2021programmable} & Yes & High, especially for long-range \\
Spin Qubits & 1/2 & 12 \cite{george202412} & Limited, typically short-range & Low, short-range focus \\
D-Wave & 1/2 & >5,000 \cite{willsch2022benchmarking} & Yes, programmable & High for Ising, less for XY \\
Cold Atoms & 9/2 & Thousands \cite{gross2017quantum} & Possible with Rydberg techniques & High for higher $S$, variable range \\
\bottomrule
\end{tabular}
\caption{Comparison of experimental platforms for simulating long-range spin XY and Ising models.}
\end{table*}

\section{Discussion and Conclusion}

In this study, we extend the Boson sampling framework to incorporate arbitrary spin-$S$ states, modeled using an Ising system with a strong transverse magnetic field. We demonstrate its experimental feasibility with trapped ions and cold atom platforms. A key finding is the quasi-linear scaling between the number of spins, $n$, and the total number of sites, $m$, which determines the quantum system’s scale. By choosing a moderate spin quantum number, such as $S = 3$, the required number of sites decreases exponentially compared to the spin-$\frac{1}{2}$ case, improving the efficiency of quantum simulations.

While the 2020 demonstration of quantum supremacy by the Chinese team was a major milestone, it had a notable limitation: it employed Gaussian Boson Sampling with a Gaussian light source. The computational hardness of Gaussian Boson Sampling is not rigorously established, as it relies on conjectures such as the Hafnian-of-Gaussians Conjecture, which are less mature than the Permanent-of-Gaussians Conjecture underpinning the proven hardness of Fock-state Boson Sampling \cite{valiant1979complexity,scheel2004permanents}. In contrast, the original Aaronson-Arkhipov paper rigorously proved the computational hardness of Fock-state Boson Sampling in the high-mode regime ($m > n^2$), where the birthday paradox does not apply. Intriguingly, a proof of the computational hardness of Fock-state Boson Sampling in the low-mode regime where $m\sim 2.1 n$ has only emerged recently \cite{bouland2023complexity}. This regime generally leads to higher collision probabilities and gives rise to the bosonic birthday paradox, thereby reducing computational hardness. However, the rigorous proof of hardness remains firmly established. This advancement opens the possibility of making Fock-state Boson Sampling feasible within the linear mode region. Nonetheless, due to experimental challenges in scaling up a strict single-photon light source, quantum supremacy via this approach has yet to be definitively achieved.

In contrast, our approach provides a practical route to achieving quantum supremacy by employing higher-spin states. The key advantage lies in the fact that the linear transformations of spin states are governed by the matrix permanent. The computational complexity of this approach is rooted in the same theoretical principles established by the Aaronson-Arkhipov proof, ensuring its robustness and validity. Furthermore, as demonstrated in the main text, a moderate spin quantum number allows for quasi-linear scaling, given by $m\sim n^{1+\epsilon}$, where $\epsilon=3/(2S)$ is inversely proportional to the spin quantum number. This scaling effectively prevents the birthday paradox, thereby ensuring computational hardness. Thus, once quantum computing platforms such as trapped ion or cold atom systems are capable of simulating long-range interactions of higher spins and scaling up to approximately hundreds of spins, quantum supremacy could be more firmly established.

For a robust demonstration of quantum supremacy in spin systems, the required number of spins depends on the specific model and interaction type. While hundreds of spins may be sufficient in some cases (e.g., 100–500 spins), achieving quantum advantage often necessitates careful optimization of system parameters, error correction, and long-range interactions.

As a final remark, some theoretical computational scientists view the Boson sampling task as fundamentally impractical, as it does not directly contribute to solving other important problems. In a sense, this perspective holds true. However, on the other hand, the experimental demonstration of a spin sampling task—one that has been rigorously proven to be computationally hard—would signify the maturity of quantum computing in the noisy intermediate-scale regime. Unlike photon-based platforms, spin sampling is performed on universal quantum computing architectures, further reinforcing its practical significance.

\section{Acknowledgments}
EJK acknowledges the support from National Yang Ming Chiao Tung University.

\begin{appendix}
\section{Birthday Paradox for Spin-$S$ Sampling}

The Birthday Paradox in statistics examines the number of people required in a group to reach a 50\% probability that at least two individuals share the same birthday \cite{sayrafiezadeh1994birthday}. The classical Birthday Paradox states that in a group of only 23 individuals, there is approximately a 50\% probability that at least two of them share the same birthday. This assumes 365 possible birthdays, with an even distribution among them. The probability that $n$ classical \textit{distinguishable} particles occupy $m$ distinct modes without collisions is
\begin{equation}\label{PNoCollision}
    P_{\text{no collision}} = \prod_{k=0}^{n-1} \left(1 - \frac{k}{m}\right).
\end{equation}
Eq.\:\eqref{PNoCollision} can be approximated as
\begin{equation}
    P_{\text{no collision}} \approx e^{-\sum_{k=1}^{n-1} \frac{k}{m}} = e^{-\frac{n(n-1)}{2m}}
\end{equation}
for large $m\gg n$. This captures the exponential decay of the probability as $n$ increases relative to $m$, while still satisfying $n \ll m$. The probability of at least one collision is given by
\begin{equation}
P_{\text{collision}} = 1 - P_{\text{no collision}} \approx 1 - e^{-\frac{n(n-1)}{2m}}.
\end{equation}
When $n(n-1)/(2m) \approx 1$, or $n \approx \sqrt{2m}$, the collision probability is about $50\%$, mirroring the birthday paradox's $n \approx \sqrt{m}$ scaling (e.g., $n \approx \sqrt{365} \approx 19$ for birthdays).

In Fock-state boson sampling \cite{aaronson2011computational, arkhipov2012bosonic, urbina2016multiparticle}, the photons are \textit{indistinguishable}, and the output distribution is determined by the permanent of submatrices of $U$, where the probability of collisions depends on the specific unitary $U$. For a random Haar unitary $U$, the output distribution of indistinguishable bosons approximates the distinguishable case in the dilute limit ($m \gg n^2$), because quantum interference effects average out. The collision probability scales similarly:
\begin{equation}
    P_{\text{collision}} \approx 1 - e^{-\frac{n^2}{2m}}.
\end{equation}
This means collisions become likely when $n \sim \sqrt{m}$, just like the classical birthday paradox. For specific unitaries, e.g., Fourier transform or highly structured interferometers, bosonic bunching can enhance or suppress collisions. For example, the Hong-Ou-Mandel effect causes two photons to bunch (land in the same mode), increasing collision probability compared to the classical case.

In the work, we inherently assume that, during the sampling task, generalized bosons do not frequently collide or reside in the same mode, thus avoiding technical difficulties in proving the hardness of Fock-state Boson sampling. For further details, \cite{arkhipov2012bosonic} provides a comprehensive calculation of the bosonic birthday paradox. Specifically, we analyze the birthday paradox within spin systems, where each spin possesses a finite spin quantum number $S$. We also restrict the highest occupation number for a single site to $2S+1$. In the following, we provide a detailed mathematical analysis of the Birthday Paradox in the context of finite spin systems. We follow Aaronson's original proof \cite{aaronson2011computational} of the Birthday paradox, adapting its mathematical framework to establish hardness results for spin sampling. Given positive integers \( m \geq n \), Aaronson's notation defines: \( \Phi_{m,n} \) as the universal set of nonnegative integer tuples \( (s_1, \dots, s_m) \) such that \( \sum_{i=1}^m s_i = n \), \( G_{m,n} \) as the subset of \( \Phi_{m,n} \) consisting of collision-free tuples, where each \( s_i \) is either 0 or 1, and \( B_{m,n} \) as the complement of \( G_{m,n} \) in \( \Phi_{m,n} \), representing tuples where at least one \( s_i \) is greater than 1.

The corresponding sets of possible outcomes for a spin-$S$ system are defined as follows
\begin{subequations}
\begin{align}
    \Phi_{m,n}^{S}& \equiv \{(s_1, . . . , s_m) \mid 2S \geq s_i \geq 0, \sum_{i=1}^m s_i = n \}, \\
    G_{m,n}^{S}& \equiv \{(s_1, . . . , s_m) \mid s_i \in \{ 0,1 \}, \sum_{i=1}^m s_i = n \}.
\end{align}
\end{subequations}
To estimate the ratio $|G_{m,n}^{S}|/|\Phi_{m,n}^{S}|$, we begin by noting that since $|\Phi_{m,n}^{S}| \leq |\Phi_{m,n}|$ and $|G_{m,n}^{S}|=|G_{m,n}|$, we immediate obtain the following lower bound
\begin{equation}\label{eq:trivial}
    \frac{|G_{m,n}^{S}|}{|\Phi_{m,n}^{S}|} \geq \frac{|G_{m,n}|}{|\Phi_{m,n}|} > 1 - \frac{n^2}{m},
\end{equation}
where the last inequality follows from Eq. (486) in the original Aaronson-Arkhipov paper \cite{aaronson2011computational}. In particular, for spin $\frac{1}{2}$, the ratio is rigorously one by definition. As a consequence of Eq.\:\eqref{eq:trivial}, the probability of the output spin distribution containing two or more photons in the same mode can be upper-bounded, yielding the same result as Eq.\:(491) in \cite{aaronson2011computational}. A tighter bound for arbitrary spin-$S$ system can be obtained, but it requires additional effort (see below).

In the following, we investigate the birthday paradox of finite spin $S$ quantitatively. We will demonstrate that Aaronson's proof for the dilute limit can be reproduced while avoiding the birthday paradox due to Eq.\:\eqref{eq:trivial}. We first recall the simple combinatorics problem:
\begin{equation}\label{Combination}
    \sum_{i=1}^m x_i = m, \,\,\, x_i \geq 0,\,\,\, i = 1,\:\cdots\:,m.
\end{equation}
The total number of solutions to Eq.\:\eqref{Combination} is given by the binomial coefficient $\binom{n+m-1}{m-1}$. This serves as a useful illustration of the technique applied in more intricate cases. Below, we employ the Hardy-Littlewood circle method \cite{vaughan2003hardy} to derive the same result. Notice that the number of solutions of Eq.\:\eqref{Combination} can be written as
\begin{align}
&\sum_{x_1,\:\dots\:,x_{m}=0}^{n}\delta_{x_{1} + \cdots + x_{m},n} \nonumber\\
=&\sum_{x_1,\:\dots\:,x_{m}=0}^{\infty}
\delta_{x_{1} + \cdots + x_{m},n}\nonumber\\
=&\sum_{x_1,\:\dots\:,x_{m}=0}^{\infty}
\oint_{|z| =a}{\frac{1}{z^{-x_{1}\ -\ \cdots\ -\ x_{m}\ +\ n\ +\ 1}}}
\,{\frac{dz}{2\pi i}}\nonumber\\
=&\oint_{|z| =a}{\frac{1}{z^{n + 1}}}\sum_{x_1,\:\dots\:,x_{m}=0}^{\infty}
{z^{x_{1}\ +\ \cdots\ +\ x_{m}}}
\,{\frac{dz}{2\pi i}}\nonumber\\
=&\oint_{|z| =a}{\frac{1}{z^{n + 1}}}\left({\frac{1}{1 - z}}\right)^{m}
\,{\frac{dz}{2\pi i} }\nonumber\\
=&\oint_{|z| =a}{\frac{1}{z^{n + 1}}}
\sum_{k = 0}^{\infty}\binom{-m}{k}(-z)^k\,{\frac{dz}{2\pi i} }\nonumber\\
=&\sum_{k = 0}^{\infty}\binom{m+k-1}{k}
\oint_{|z| =a}{\frac{1}{z^{n - k + 1}}}\,{\frac{dz}{2\pi i} }\nonumber\\
=&\sum_{k = 0}^{\infty}\binom{m+k-1}{k}\delta_{k,n}
=\binom{m+n-1}{n}.
\end{align}
In the analysis above, we have utilized the contour integral property $\oint_{|z| =|a|}\frac{1}{2\pi i z^n}=\delta_{n,-1}$ \cite{stein2010complex} to determine the number of solutions, where the integration is performed over a circular contour centered at the origin with radius $a$. In the last second step, we have used the identity $\binom{m+k-1}{k}(-1)^k=\binom{-m}{k}.$

Building upon the previous example, we now extend our method to the finite spin case, employing the same technique. Our objective is to determine the number of solutions to the following combinatorial problem
\begin{equation}
    \sum_{i=1}^m x_ i= n, \,\,\, 0 \leq x_i \leq 2S,\,\,\, i \in \{1,\:\cdots\:,m\}.
\end{equation}
If $n \leq 2S$, we observe that the result remains unchanged. Therefore, let us now assume $n > 2S$ and proceed by applying the same Hardy-Littlewood circle method
\begin{align}
&\sum_{x_{1},\:\cdots\:,x_{m} = 0}^{2S}\delta_{x_{1} + \cdots + x_{m},n}\nonumber\\
=&\sum_{x_{1},\:\cdots\:,x_{m} = 0}^{2S}
\oint_{|z| =a}{\frac{1}{z^{-x_{1}\ -\ \cdots\ -\ x_{m}\ +\ n\ +\ 1}}}
\,{\frac{dz}{2\pi i}}\nonumber\\
=&\oint_{|z| =a}{\frac{1}{z^{n + 1}}}\left(\frac{1-z^{2S+1}}{1 - z}\right)^{m}
\,{\frac{dz}{2\pi i} }\nonumber\\
=&\oint_{|z| =a}{\frac{1}{z^{n + 1}}}
\sum_{l = 0}^{m}\binom{m}{l}(-1)^l z^{(2S+1)l}
\sum_{k = 0}^{\infty}\binom{-m}{k}(-z)^k\,{\frac{dz}{2\pi i} }\nonumber\\
=&\sum_{l = 0}^{m}\sum_{k = 0}^{\infty}\binom{m+k-1}{k}\binom{m}{l}
\oint_{|z| =a}{\frac{(-1)^l}{z^{n - k- (2S+1)l+1}}}\,{\frac{dz}{2\pi i} }\nonumber\\
=&\sum_{l = 0}^{m} \binom{m +n-(2S+1)l - 1}{n-(2S+1)l} \binom{m}{l}(-1)^l.
\end{align}
Thus the number of the universal set $\Phi_{m,n}^{S}$ of nonnegative integer tuples for a spin-$S$ system is
\begin{equation}
   |\Phi_{m,n}^{S}| = \sum_{l = 0}^{\lfloor \frac{n-1}{2S+1} \rfloor}{m +n-(2S+1)l - 1 \choose m-1}{m \choose l}(-1)^l. 
\end{equation}
We now assume that the number of spins follows a power-law scaling relation with the number of sides, given by $m= O(n^a)$, where $a > 2$ is a constant. For sufficiently large $n$, the following relation holds
\begin{equation}
\frac{m^l}{l!}\frac{{m +n- (2S+1)l-1 \choose n-(2S+1)l}}{{m+n-1 \choose n}}=\frac{n^{l-2 (a-1) l S}}{l!}(1+O(\frac{1}{n})).
\end{equation}
Since the ratio decreases as $l$ increases, higher-order corrections become less significant, making the $l=1$ term the primary adjustment to the leading order approximation
\begin{equation*}
  |\Phi_{m,n}^{S}|= {m+n-1 \choose n}[1-n^{1-2 (a-1) S}+O(n^{1-4(a-1) S}) ],
\end{equation*}
which immediately yields 
\begin{equation}\label{eq:short}
\frac{|\Phi_{m,n}^{S}|}{|\Phi_{m,n}|}= 1-\frac{n^{2S+1}}{m^{2S}}+O(\frac{n^{4S+1}}{m^{4S}}).
\end{equation}
For $S=\frac{1}{2},$ we obtain the standard birthday paradox, whereas for $S \geq 1,$ we have an asymptotic bound for large $n$
\begin{align}
\label{eq:spin-birth}
\frac{|G_{m,n}^{S}|}{|\Phi_{m,n}^S|} &= \frac{|G_{m,n}|}{|\Phi_{m,n}|}\left(1+O\left(\frac{n^{1+2S}}{m^{2S}}\right)\right) \nonumber\\
    &> 1 - \frac{n^2}{m} + O\left(\frac{n^{1+2S}}{m^{2S}}\right),
\end{align}
where in the last step we have used \cite{aaronson2011computational}
\begin{equation}\label{eq:baseline}
    \frac{|G_{m,n}|}{|\Phi_{m,n}|} = \frac{{m \choose n}}{{m+n-1 \choose m-1}} > 1 - \frac{n^2}{m} = e^{-\frac{n^2}{m}} + O\left(\frac{n^4}{m^2}\right).
\end{equation}
Based on a similar reasoning outlined in \cite{aaronson2011computational} and the spin $1/2$ proof presented in \cite{peropadre2017equivalence}, the reduction of the bound $\|P\ket{\varepsilon(\tau)}\|^2$ leads to a combinatorial counting problem
\begin{align}
\|P\ket{\varepsilon(\tau)}\|^2 \leq 1-\frac{|\Phi_{m,n}^{S}|}{|\Phi_{m,n}|}.  
\end{align}
Using Equation.~\eqref{eq:short}, this result highlights the crucial spin scaling behavior described in
Eq.~\eqref{SM_eq:paradox_bound}.

\section{Permanent, Hafnian and Torontonian in Boson Sampling}
Fock-state Boson sampling and Gaussian Boson sampling are quantum computational tasks designed to demonstrate quantum supremacy, or quantum advantage, using experimentally feasible quantum machines \cite{aaronson2011computational, quesada2018gaussian}. These tasks involve processing input states—either Fock states or Gaussian states of bosons—to highlight the computational power of quantum systems beyond classical capabilities. The computational complexity of these tasks relies on fundamental matrix functions: the Permanent for Boson Sampling, and the Hafnian or Torontonian for Gaussian Boson Sampling. This appendix supplements these matrix functions for general readers, complementing the main text.

\subsection{Permanents in Fock-state Boson Sampling}
The permanent of an \( n \times n \) matrix \( A = (a_{ij}) \) is defined as \cite{papadimitrioucomputational}
\begin{equation}
\text{perm}(A) = \sum_{\sigma \in S_n} \prod_{i=1}^n a_{i,\sigma(i)},
\end{equation}
where \( S_n \) is the set of all permutations of \( \{1, 2, \dots, n\} \). Unlike the determinant, the permanent sums all terms with positive coefficients. For a 2$\times$2 matrix $ A = \begin{pmatrix} a & b \\ c & d \end{pmatrix} $, $\text{perm}(A) = ad + bc.$

In Boson sampling, a linear optical network with \( m \) modes is described by an \( m \times m \) unitary matrix \( U \). The input is \( n \) single photons in \( m \) modes (\( m \gg n \)), and the output is a photon configuration \( S = (s_1, s_2, \dots, s_m) \), with \( \sum s_i = n \). The probability of observing \( S \) given input \( T = (t_1, t_2, \dots, t_m) \), with $\sum t_i=n$, is \cite{aaronson2011computational}
\begin{equation}
P(S|T) = \frac{|\text{perm}(U[S|T])|^2}{\prod_{i=1}^m s_i! \prod_{j=1}^m t_j!},
\end{equation}
where \( U[S|T] \) is an \( n \times n \) sub-matrix of \( U \), constructed by selecting rows and columns based on \( S \) and \( T \), accounting for photon multiplicities.

To illustrate the construction of the matrix \( U[S|T] \) and the computation of its permanent, we consider a linear optical network with \( m = 4 \) modes, described by the unitary matrix:
\begin{equation}
U = \frac{1}{2} \begin{pmatrix}
    1 & 1 & 1 & 1 \\
    1 & i & -1 & -i \\
    1 & -1 & 1 & -1 \\
    1 & -i & -1 & i
\end{pmatrix}.
\end{equation}
This matrix satisfies \( U^\dagger U = I \), and its elements are given by \( u_{ij} = \frac{1}{2} \cdot \omega_{ij} \), where \( \omega_{ij} \) belongs to the set \( \{1, i, -1, -i\} \). The input state is defined as $|s\rangle = |2, 1, 0, 0\rangle$, which represents 2 photons in mode 1, 1 photon in mode 2, and none in modes 3 and 4—totaling 3 photons. The output state is $|t\rangle = |1, 1, 1, 0\rangle$, meaning 1 photon occupies modes 1, 2, and 3, with none in mode 4, preserving the total photon count at 3. To construct \( U[S|T] \), we repeat the \( i \)-th row of \( U \) \( s_i \) times and the \( j \)-th column \( t_j \) times. Once assembled, we compute its permanent to determine the transition probability between \( |s\rangle \) and \( |t\rangle \).

The matrix \( U[S|T] \) is a \( 3 \times 3 \) matrix (since \( n = 3 \)) constructed as follows
\begin{itemize}
    \item Rows (from input \( |s\rangle \)):
    \begin{itemize}
        \item 1st row of \( U \), \( s_1 = 2 \) times.
        \item 2nd row of \( U \), \( s_2 = 1 \) time.
        \item Rows 3 and 4 are excluded (\( s_3 = 0, s_4 = 0 \)).
    \end{itemize}
    \item Columns (from output \( |t\rangle \)):
    \begin{itemize}
        \item 1st column of \( U \), \( t_1 = 1 \) time.
        \item 2nd column of \( U \), \( t_2 = 1 \) time.
        \item 3rd column of \( U \), \( t_3 = 1 \) time.
        \item Column 4 is excluded (\( t_4 = 0 \)).
    \end{itemize}
\end{itemize}
The rows of \( U \) are row 1: $( \frac{1}{2}, \frac{1}{2}, \frac{1}{2}, \frac{1}{2} )$, and row 2: $( \frac{1}{2}, \frac{i}{2}, \frac{-1}{2}, \frac{-i}{2} )$. The required columns of \( U \) are column 1: \( ( \frac{1}{2}, \frac{1}{2}, \frac{1}{2}, \frac{1}{2} )^T \), column 2: \( ( \frac{1}{2}, \frac{i}{2}, \frac{-1}{2}, \frac{-i}{2} )^T \), and column 3: \( ( \frac{1}{2}, \frac{-1}{2}, \frac{1}{2}, \frac{-1}{2} )^T \). Thus, \( U[S|T] \) is determined by
\begin{equation}
U[S|T] = \frac{1}{2} \begin{pmatrix}
1 & 1 & 1 \\
1 & 1 & 1 \\
1 & i & -1
\end{pmatrix}.
\end{equation}
Since the permanent of a $3\times 3$ matrix with elements \( a_{ij} \) is given by $\sum_{\sigma \in S_3} a_{1,\sigma(1)} a_{2,\sigma(2)} a_{3,\sigma(3)}$, where \( S_3 \) is the set of permutations of \( \{1, 2, 3\} \): \( (1,2,3), (1,3,2), (2,1,3), (2,3,1), (3,1,2), (3,2,1) \). For each permutation, we can compute the product for \( U[S|T] \) 
\begin{itemize}
    \item \( \sigma = (1,2,3) \): $a_{1,1} a_{2,2} a_{3,3} = \frac{1}{2} \cdot \frac{1}{2} \cdot \frac{-1}{2} = \frac{-1}{8}.$
    \item \( \sigma = (1,3,2) \): $a_{1,1} a_{2,3} a_{3,2} = \frac{1}{2} \cdot \frac{1}{2} \cdot \frac{i}{2} = \frac{i}{8}.$
    \item \( \sigma = (2,1,3) \): $a_{1,2} a_{2,1} a_{3,3} = \frac{1}{2} \cdot \frac{1}{2} \cdot \frac{-1}{2} = \frac{-1}{8}.$
    \item \( \sigma = (2,3,1) \): $a_{1,2} a_{2,3} a_{3,1} = \frac{1}{2} \cdot \frac{1}{2} \cdot \frac{1}{2} = \frac{1}{8}.$
    \item \( \sigma = (3,1,2) \): $a_{1,3} a_{2,1} a_{3,2} = \frac{1}{2} \cdot \frac{1}{2} \cdot \frac{i}{2} = \frac{i}{8}.$
    \item \( \sigma = (3,2,1) \): $a_{1,3} a_{2,2} a_{3,1} = \frac{1}{2} \cdot \frac{1}{2} \cdot \frac{1}{2} = \frac{1}{8}.$
\end{itemize}
As such, the permanent of \( U[S|T] \) is determined to be
\begin{equation}
\text{perm}(U[S|T]) = \frac{i}{4}.
\end{equation}
Hence, the probability for the output photon configuration is determined by
\begin{align}
    P(|s\rangle | |t\rangle) &= \frac{|\text{perm}(U[s|t])|^2}{\prod_i s_i! \prod_j t_j!}\nonumber\\
    &=\frac{\frac{1}{16}}{2} = \frac{1}{32}.
\end{align}
This example, with 3 photons and 4 modes, is relatively small. Scaling to 20–30 photons renders classical simulation infeasible due to the exponential growth in permanent computation complexity, highlighting boson sampling’s potential for demonstrating quantum computational advantage. In general, computing the permanent is \#P-hard \cite{valiant1979complexity}, as it involves \( n! \) terms. Unlike the determinant, which can be efficiently computed using algorithms such as Gaussian elimination, no comparable simplification method exists for the permanent. This hardness underpins the quantum advantage of Boson Sampling.

\subsection{Hafnians in Gaussian Boson Sampling}

The Hafnian of a \( 2n \times 2n \) symmetric matrix \( A = (a_{ij}) \) is \cite{caianiello1953combinatorics}
\[
\text{haf}(A) = \sum_{\text{perfect matchings } M} \prod_{(i,j) \in M} a_{ij},
\]
where the sum is over all perfect matchings of a complete graph on \( 2n \) vertices \cite{barvinok2016combinatorics}. A perfect matching pairs all vertices into \( n \) edges. For example, for a $4\times 4$ symmetric matrix, the Hafnian is determined by:
\[
\text{haf}(A) = a_{12}a_{34} + a_{13}a_{24} + a_{14}a_{23}.
\]
In Gaussian Boson sampling, the input consists of Gaussian states, e.g., squeezed vacuum states, in \( m \) modes, transformed by a linear optical network. The output is a photon configuration \( S = (s_1, s_2, \dots, s_m) \). The probability is:
\[
P(S) = \frac{1}{\sqrt{\det(Q)}} \cdot \frac{|\text{haf}(A_S)|^2}{\prod_{i=1}^m s_i!},
\]
where \( A_S \) is a \( 2n \times 2n \) symmetric matrix (\( n = \sum_{i=1}^m s_i \)) derived from the covariance matrix of the transformed Gaussian state, adjusted for the photon configuration \( S \). \( \text{haf}(A_S) \) is the Hafnian of \( A_S \), representing the quantum amplitude for the configuration \( S \). \( \prod_{i=1}^m s_i! \) accounts for the indistinguishability of photons in each mode. \( Q \) is a matrix related to the covariance matrix of the Gaussian state, ensuring proper normalization. \( \sqrt{\det(Q)} \) is part of the normalization factor, ensuring that the probabilities sum to 1 over all possible configurations: $\sum_S P(S) = 1$.

The Hafnian is \#P-hard to compute \cite{barvinok2016combinatorics}, involving $(2n-1)!!$ terms. Counting perfect matchings in a general non-bipartite graph is known to be \#P-complete, as shown in computational complexity theory via reductions from other hard counting problems such as counting matchings or satisfiability \cite{valiant1979complexity}. In this context, the Hafnian generalizes the perfect matching counting problem to weighted matchings where matrices have arbitrary entries. As a result, computing the Hafnian exactly remains \#P-hard even for real or complex matrices \cite{barvinok2016combinatorics}, which is crucial in Gaussian Boson Sampling. This inherent difficulty underpins the classical intractability of simulating Gaussian Boson Sampling.

\subsection{Torontonians in Gaussian Boson sampling}

The Torontonian is defined as \cite{quesada2018gaussian}
\begin{equation}
\text{tor}(A) = \sum_{S \subseteq \{1, \dots, m\}} (-1)^{m - |S|} \det(A_{\bar{S},\bar{S}}),
\end{equation}
where
\begin{itemize}
    \item \( S \) is a subset of the \( m \) modes.
    \item \( \bar{S} \) is the complement of \( S \).
    \item \( A_{\bar{S},\bar{S}} \) is the submatrix of \( A \) with rows and columns indexed by \( \bar{S} \).
    \item The sum is over all \( 2^m \) subsets \( S \).
\end{itemize}

In Gaussian Boson sampling, a Gaussian state, e.g., squeezed vacuum states, in \( m \) modes is processed through a linear optical network, and the output photon distribution is sampled. Threshold detectors simplify the measurement compared to PNR detectors by recording only whether a mode contains at least one photon. Threshold detectors were not used in Jiuzhang, which relied on PNR detectors.

\begin{table}[tbp]
\centering
\begin{tabular}{|l|c|c|}
\toprule
\textbf{} & \textbf{Boson Sampling} & \textbf{Gaussian Boson Sampling} \\
\midrule
\textbf{Matrix Function} & Permanent & Hafnian or Torontonian \\
\textbf{Matrix Type} & \( n \times n \) complex & \( 2n \times 2n \) symmetric \\
\textbf{Combinatorial Sum} & Permutations & Perfect matchings or subsets \\
\textbf{Complexity} & \#P-hard & \#P-hard \\
\textbf{Physical Context} & Single photons & Gaussian states \\
\bottomrule
\end{tabular}
\caption{Comparison of mathematical functions used in Boson Sampling and Gaussian Boson Sampling.}
\end{table}

The output is a binary vector \( C = (c_1, \dots, c_m) \), where:
\begin{itemize}
    \item \( c_i = 1 \) if mode \( i \) has at least one photon (a ``click'').
    \item \( c_i = 0 \) if mode \( i \) has no photons (``no click'').
\end{itemize}
For example, \( C = (1, 0, 1, 0) \) indicates clicks in modes 1 and 3, and no clicks in modes 2 and 4. There are \( 2^m \) possible click patterns. The probability of observing a click pattern \( C \) is proportional to the Torontonian of a matrix \( A_C \): $P(C) \propto \text{tor}(A_C)$, where \( A_C \) is a \( 2m \times 2m \) matrix derived from the covariance matrix of the output Gaussian state, adjusted for the click pattern \( C \). The proportionality reflects the need for a normalization constant:
\begin{equation}
    C = \left( \sum_C \text{tor}(A_C) \right)^{-1},
\end{equation}
to ensure \( \sum_C P(C) = 1 \). The \#P-hardness of the Torontonian was established in quantum optics literature \cite{quesada2018gaussian}. It can be shown to be \#P-hard by reducing a known \#P-complete problem, such as counting perfect matchings or computing the permanent, to the Torontonian \cite{barvinok2016combinatorics}.

To conclude, the permanent, Hafnian, and Torontonian are \#P-hard matrix functions central to the quantum advantage of Boson Sampling and Gaussian Boson Sampling. The permanent captures single-photon interference, while the Hafnian and Torontonian reflect correlations in Gaussian states, with the latter used for threshold detection.
\end{appendix}

\bibliographystyle{unsrt}
\bibliography{sample}

\onecolumngrid

\end{document}